\documentclass[10pt, letterpaper]{IEEEtran}  

\usepackage{graphics} 
\usepackage{amsmath} 
\usepackage{mathtools}
\usepackage{amssymb}  
\usepackage{color}
\usepackage[colorlinks]{hyperref}
\usepackage[ruled,vlined,linesnumbered]{algorithm2e} 
\usepackage{subfigure}
\usepackage{booktabs}
\usepackage{caption}

\graphicspath{{./figures/}}

\newcommand{\sysname}{BD-SAS\xspace}

\title{Decentralized Spectrum Access System: Vision, Challenges, and a Blockchain Solution}

\author{
    \IEEEauthorblockN{
        {Yang Xiao}\IEEEauthorrefmark{1},
        {Shanghao Shi}\IEEEauthorrefmark{1},
        {Wenjing Lou}\IEEEauthorrefmark{1},
        {Chonggang Wang}\IEEEauthorrefmark{2},
        {Xu Li}\IEEEauthorrefmark{2},\\
        {Ning Zhang}\IEEEauthorrefmark{3},
        {Y. Thomas Hou}\IEEEauthorrefmark{1},
        {Jeffrey H. Reed}\IEEEauthorrefmark{1}
      } \\
    \IEEEauthorrefmark{1}\IEEEauthorblockA{Virginia Polytechnic Institute and State University, Blacksburg, VA, USA}\\
    \IEEEauthorrefmark{2}\IEEEauthorblockA{InterDigital, Princeton, NJ, USA}\\
    \IEEEauthorrefmark{3}\IEEEauthorblockA{Washington University in St. Louis, MO, USA}\\
    \thanks{*A version of this work has been accepted by IEEE Wireless Communications for publication.}
    \thanks{\copyright 2021 IEEE. Personal use of this material is permitted. Permission from IEEE must be obtained for all other uses, including reprinting/republishing this material for advertising or promotional purposes, collecting new collected works for resale or redistribution to servers or lists, or reuse of any copyrighted component of this work in other works.}
    }

\begin{document}


\maketitle

\begin{abstract}
Spectrum access system (SAS) is widely considered the de facto solution to coordinating dynamic spectrum sharing (DSS) and protecting incumbent users.
The current SAS paradigm prescribed by the FCC for the CBRS band and standardized by the WInnForum follows a centralized service model in that a spectrum user subscribes to a SAS server for spectrum allocation service. This model, however, neither tolerates SAS server failures (crash or Byzantine) nor resists dishonest SAS administrators, leading to serious concerns on SAS system reliability and trustworthiness. 
This is especially concerning for the evolving DSS landscape where an increasing number of SAS service providers and heterogeneous user requirements are coming up. 
To address these challenges, we propose a novel blockchain-based decentralized SAS architecture called \sysname that provides SAS services securely and efficiently, without relying on the trust of each individual SAS server for the overall system trustworthiness.
In \sysname, a global blockchain (G-Chain) is used for spectrum regulatory compliance while smart contract-enabled local blockchains (L-Chains) are instantiated in individual spectrum zones for automating spectrum access assignment per user request.
We hope our vision of a decentralized SAS, the \sysname architecture, and discussion on future challenges can open up a new direction towards reliable spectrum management in a decentralized manner. 
\end{abstract}
\begin{IEEEkeywords}
Spectrum sharing, spectrum access system, blockchain
\end{IEEEkeywords}


\section{Introduction}
Spectrum is the single most important resource for wireless communication and sensing. With the advent of 5G and the proliferation of wireless devices, regulators---mainly the Federal Communications Commission (FCC) and the National Telecommunications and Information Administration (NTIA) in the US---have opened up previously licensed bands (e.g., sub-GHz TVWS and 3.5 GHz CBRS) and unlicensed bands (e.g., 6 GHz to mmWave) for civilian use on a sharing basis. 
To protect the access rights of incumbent users and ensure fairness in the dynamic spectrum sharing (DSS) process, the FCC has stipulated the need for a spectrum access system (SAS) for spectrum management in its rulings on the CBRS band \cite{fcc2015rule}. The current standardization effort on SAS for CBRS \cite{WIForum-SAS,WIForum-CBSD}, led by the Wireless Innovation Forum (WInnForum), delegates the tasks of incumbents protection, user device registration, and dynamic spectrum allocation to individual SAS servers. Each SAS server is proprietary to a commercial entity known as SAS administrator (e.g., Google, Federated Wireless, CommScope). Each SAS server maintains a spectrum database and generates spectrum access assignment (including allocated channel and time) in response to users' spectrum access request, resembling a server-client model. SAS servers may communicate with each other for service updates. 

While the centralized SAS paradigm is an effective solution for DSS, it assumes all SAS administrators are trustworthy and each SAS server is able to make policy-abiding assignment decisions. This assumption, however, may not hold in the evolving threat landscape of spectrum sharing. It is envisioned that there will be a large number of SAS servers providing a diverse set of specialized spectrum management services based on the heterogeneous spectrum regulations.
Without the ability to ensure spectrum policies are strictly enforced in these individual SAS servers, a faulty or compromised SAS server can provide erroneous or maliciously crafted assignments to spectrum users which could wreak havoc to the users' operation and the wireless network as a whole. 
%
Meanwhile, reliable access to spectrum has already become a daily necessity for spectrum users; those who rely on the SAS for dynamic access assignment are no exceptions.
It is crucial that the users are able to utilize the spectrum even if individual SAS servers are compromised. 
In this regard, a decentralized, collectively managed SAS architecture is more desirable in that spectrum access assignments are finalized via fault-tolerant consensus across stakeholders. All stakeholders, spectrum users and SAS servers alike, are encouraged to participate in the management process, enforce the spectrum regulations, and contribute to the system's robustness.
Blockchain recently emerged as a secure-by-design technology that powers fully decentralized payment networks. With the cryptography-hardened transactional model and consensus-based validation mechanism, blockchain enables trusted transaction processing and ledger keeping among mutually distrustful participants, even if a certain portion of them behave maliciously \cite{xiao2020survey}. 
The decentralized zero-trust nature of blockchain has also been considered by the FCC as a possible paradigm to enable spectrum sharing in the future \cite{rosenworcel2018fcc}.
In this paper, we first establish the functional and security requirements of decentralized SAS and propose a blockchain-based solution called \sysname to fill the gap. \sysname is comprised of two layers of blockchain networks: the G-Chain at the global scale and L-Chains for local spectrum zones. The G-Chain participants include SAS servers and regulator nodes, who maintain a unified blockchain ledger on spectrum regulations and digests of local SAS service states. An L-Chain is dedicated to spectrum access management for a specific geographical zone (e.g., a county in the CBRS case) and maintained by SAS servers who serve that zone and local stable users called witnesses. To enable automated spectrum access assignment, a spectrum access contract encoding an allocation function is established on the L-Chain. The function is invoked at a spectrum user's request and outputs an assignment decision indicating the allocated channels and time to use. Essentially, the correctness of the function execution is safeguarded by L-Chain's underlying consensus mechanism. Assignment results are finalized in the L-Chain ledger and open for audit.

While \sysname is designed for the 3.5GHz CBRS band, the design can be easily extended to spectrum access management across different bands and collaborative spectrum sharing in general (i.e., in contrast to fully opportunistic spectrum sharing). CBRS is chosen as the case study because it has been the forefront for SAS standardization and commercial deployment. \sysname realizes the basic SAS functionalities specified in WInnForum's standardization documents \cite{WIForum-SAS,WIForum-CBSD} and is backward-compatible with the existing CBRS ecosystem.
We provide an instantiation of \sysname' L-Chain using the Hyperledger Fabric blockchain platform \cite{androulaki2018hyperledger}. Performance evaluation on an emulated L-Chain network demonstrates the feasibility of finalizing a spectrum access assignment
gracefully within the tight Heartbeat interval specified in the existing centralized SAS.

\section{Motivation for Decentralized SAS and Prospect of Blockchain Solutions}

\subsection{The Current SAS Paradigm}
\label{subsec:current-sas}
Spectrum access system (SAS) was coined by the FCC in 2012 and later formalized in a 2015 ruling on the CBRS band \cite{fcc2015rule}. In the CBRS band, spectrum users fall into three tiers: Incumbent Access (IA) tier including federal users and fixed satellite service earth stations, Priority Access (PA) tier including users who obtain Priority Access Licenses (PALs) on a county-by-county basis through competitive bidding, and General Authorized Access (GAA) tier including a broad range of daily users (e.g., private LTE network, industrial IoT, campus hot spots) who access the band in a flexible manner. In commercial deployment, PA and GAA users utilize the CBRS band via fixed stations known as citizens broadband radio service devices (CBSDs).
Under this tiered model, the SAS is designated by the FCC for managing the shared access to the CBRS band while protecting the preemptive right of IA users by acting on notifications from environmental sensing capability (ESC) entities.

Since 2016, the WInnForum has been leading the CBRS standardization effort, including specifications on the inter-SAS communication \cite{WIForum-SAS} and SAS-CBSD interface \cite{WIForum-CBSD}. Each SAS server, which is proprietary to a SAS administrator, maintains a database on the local spectrum availability and receives CBSD registrations. The spectrum assignment process follows a server-client model and consists of two main procedures---``Grant'' and ``Heartbeat'' \cite{WIForum-CBSD}. When the SAS server receives a Grant request from a registered CBSD, it responds with a Grant assignment which specifies the operational parameters including assigned channel range and expiration time, and Heartbeat interval. Then the CBSD needs to send Heartbeat requests to the SAS server periodically as proof of liveness, and receives Heartbeat responses, each of which authorizes the CBSD to transmit in the granted channels for the next Heartbeat interval. 
To maximize spectrum utilization, the Heartbeat interval can be as tight as 30 seconds \cite{google2020sasresponse} and a CBSD needs to stop radio transmission within 60 seconds after its Grant expires or gets relinquished \cite{fcc2015rule}.
To facilitate coordination across different SAS servers, SAS servers may communicate with each other and synchronize service state and CBSD records.

\subsection{Decentralized SAS: A Trustworthy, Fault-tolerant Model} 
\label{subsec:motive-decentsas}

The current SAS paradigm, though conceptually simple, faces several  challenges in terms of service reliability. 
First, spectrum users rely on one selected SAS server for spectrum access assignments. A malfunctioning or compromised SAS server can be devastating to effective spectrum utilization by users.
Second, SAS servers need to communicate with each other and synchronize service records, including CBSD registration, change in PAL Protection Area (PPA), and CBSD coordination events \cite{WIForum-SAS}. SAS servers under malicious administrators may disseminate false or tampered records in the peer network which could sabotage the operation of other SAS servers.
As more SAS administrators and servers join the ecosystem to provide spectrum management services, we must consider the possibility that one or more of them may collude to jeopardize the system's quality of service for their malicious goals. 

To address the fundamental challenge on SAS ``trust'' and security implications, we consider a \emph{decentralized SAS} ideal for providing trustworthy and fault-tolerant spectrum management service which does not need users to trust individual SAS servers or SAS servers to trust each other. Specifically, we identify four requirements for a decentralized SAS:
\begin{itemize}
    \item \textbf{Trustworthiness~} A user should receive regulation-compliant and fair spectrum access assignments from the system. The assignment process and records should be transparent and available for user/regulator auditing.
    \item \textbf{Fault Tolerance~} 
    The trustworthiness requirement holds even when a certain minority portion of SAS servers arbitrarily deviate from their normal routines, be it malfunctioning or under malicious attacks.
    \item \textbf{Responsiveness~} The processing of a user's spectrum access request and finalization of an assignment should be swift, on par with the existing service requirements (currently on the seconds time scale for CBRS). 
    \item \textbf{Scalability~} The SAS should accommodate the increasing variety and quantity of SAS administrators/servers and spectrum users, without compromising its fault tolerance capability and responsiveness.
\end{itemize}

The trustworthiness and fault tolerance requirements essentially set apart the decentralized SAS model from the existing centralized SAS paradigm that requires every CBSD operator to fully trust the SAS server it subscribes to. Adversarial influence on an individual SAS server will not affect a CBSD's operation. The responsiveness and scalability requirements are aimed to limit the performance impact of decentralization, which is usually associated with increased communication and computation redundancy.

\subsection{Prospect of Blockchain Solutions to Decentralized SAS}
\label{subsec:prospect-blockchain}

The blockchain technology becomes known for constructing decentralized payment network among mutually distrustful participants. Secured by a consensus mechanism, the shared blockchain ledger becomes a validated and irreversible record of the network's transactions when the majority of the network's voting power (e.g., hashing power, stake value, authorized identities) is controlled by honest participants. Aside from cryptocurrency, blockchain has enabled a wide range of decentralized applications (DApps) that could previously run only with a trusted intermediary.
%
Smart contract, one of the most popular DApps, realizes complex multiparty business logic and sees wide adoption in novel business scenarios, such as decentralized finance, digital identity, and supply chain management.
The decentralized and zero-trust nature, consensus-based security, transparency, and irreversible ledger make blockchain a promising technology for enabling various dynamic spectrum sharing scenarios for 5G and beyond, as is envisioned in recent surveys \cite{weiss2019application,shi2021challenges}. 
And crucially, blockchain smart contract provides an automated computing platform for realizing spectrum access assignment (e.g., calculating an assignment based on a request and an interference model) and enforcing spectrum regulations.

Prior work has explored building a blockchain-based SAS. A digital-token-based spectrum access platform is proposed in \cite{ariyarathna2019dynamic} wherein a smart contract system is used by primary spectrum users as a trusted third-party service for advertising and leasing spectral tokens to secondary users. 
In \cite{grissa2019trustsas}, a hierarchical blockchain framework called TrustSAS is formulated to enable efficient and privacy-preserving spectrum sharing among secondary users. Local blockchain networks are established among secondary users for spectrum query aggregation and response distribution while a global blockchain is used for general policies and records keeping. A blockchain-enhanced spectrum sharing system is proposed for the CBRS band \cite{zhang2020blockchain}. The PAL users are responsible for establishing local blockchain networks which help a central regulator reduce its workload in spectrum sharing coordination. 
In contrast to the decentralized SAS objective, these proposals generally assume absolute trust on either a third-party contract platform \cite{ariyarathna2019dynamic} or an authoritative SAS server \cite{grissa2019trustsas,zhang2020blockchain} and do not consider the security impact of malfunctioning or malicious SAS servers. 

\textbf{Towards a Practical Blockchain-based Solution~} Based on the discussion above, we desire a blockchain solution that fully decentralizes its decision process in spectrum sharing and also satisfies the four requirements of the decentralized SAS model. From the practical point of view, we should anticipate the performance impact of blockchain consensus.
Consensus protocols used by public/permissionless blockchains, such as cryptocurrencies, tend to suffer from highly wasteful computation, low transaction throughput, and high confirmation latency in order to attain consensus security under a zero-trust and pseudonymous participation model. 
In comparison, access-controlled/permissioned blockchains usually have stable participants with revealed identities and good inter-connectivity, which allow adoption of efficient consensus protocols that can use multicast communication and yield much better throughput and latency performance. 
A blockchain-based solution should also be backward compatible with the existing SAS, as has been standardized by the WInnForum, by inheriting its infrastructure and coexisting with its server-client model.

\section{\sysname: Blockchain-based Decentralized Spectrum Access System}

\subsection{\sysname Overview}

We present a blockchain-based decentralized SAS architecture dubbed \sysname on top of the existing CBRS infrastructure. We define four types of participants:
\begin{itemize}
    \item \textbf{Regulator} is a government entity that publishes regulative information about spectrum usage, such as protected federally owned spectrum zones, priority access licensees, etc. Examples of regulators include the FCC and the NTIA in the US. 
    \item \textbf{SAS administrator} is a licensed corporate-level entity, such as Google, Federated Wireless, and CommScope, who provides regulation-compliant commercial SAS services through its proprietary SAS servers.
    \item \textbf{SAS server} also known as SAS implementation provides registration and spectrum access assignment services to CBSD clients. It is managed by a SAS administrator and may not physically reside in its serviced spectrum zones. 
    \item \textbf{CBSD client} is a normal CBSD resembling a physical spectrum user and relies on the SAS for spectrum access. 
    \item \textbf{CBSD witness} is a CBSD who participates in local spectrum management alongside SAS servers and other CBSD witnesses. For each spectrum zone, CBSD witness candidates include all PAL-tier CBSDs and stable GAA-tier CBSDs, or to be determined on a reputation basis.
\end{itemize}

\begin{figure*}
    \centering
    \includegraphics[width=0.92\textwidth]{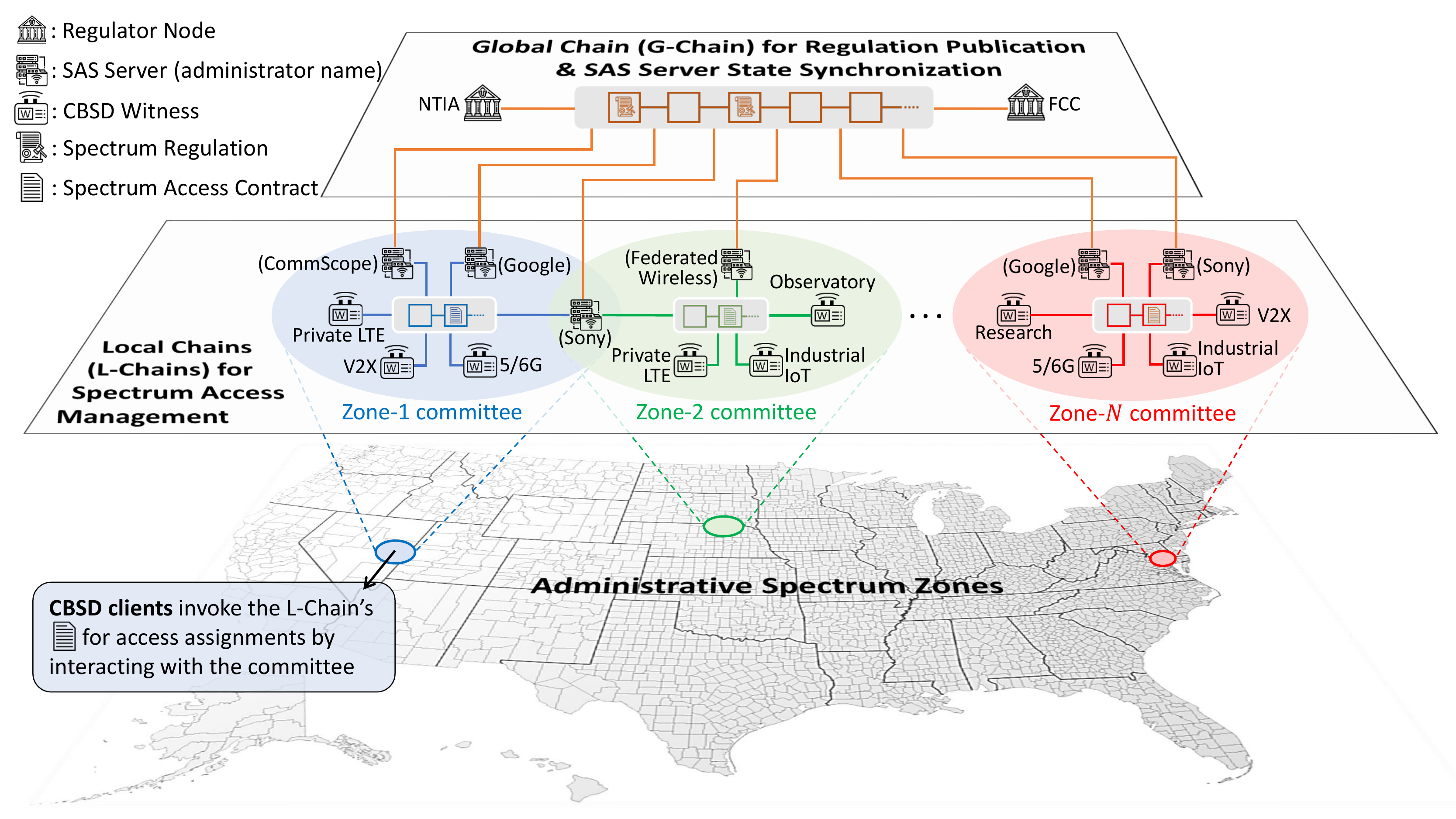}
    \caption{An illustrative example of \sysname architecture for the CBRS band.}
    \label{fig:sys-arch}
\end{figure*}

The \sysname architecture and functionalities are illustrated in Fig. \ref{fig:sys-arch}. 
\sysname is composed of two layers of blockchain networks: a single Global Chain network (G-Chain) for regulation compliance and SAS state synchronization and zone-specific Local Chain networks (L-Chains) for spectrum access assignment.
The G-Chain is curated by regulator nodes and SAS servers. Regulator nodes publish spectrum regulations onto G-Chain which will be observed by SAS servers. SAS servers also synchronize their service states through G-Chain transactions.
An L-Chain is curated by its local committee, consisting of the SAS servers that provide spectrum access assignment service to that zone and local CBSD witnesses. 
Ordinary CBSDs act as clients to the L-Chain local committee.
While a SAS server is under the direct management of its administrator, it exerts a high level of autonomy when participating in a local spectrum committee.
The information exchange between the G-Chain and L-Chains is performed by SAS servers. Selected local information such as CBSD registrations and coordination events, are uploaded to G-Chain by SAS servers for service state synchronization.
Compared to the existing SAS architecture, \sysname features a consensus-based fault-tolerant decision process at both the global level (for inter-SAS server information sharing) and local level (for generating spectrum access assignment). The G-Chain and L-Chain ledgers constitute an irreversible operation history that can be audited by an authorized \sysname participant.

\subsection{G-Chain}

The G-Chain is established among regulator nodes and SAS servers at the global scale and fulfills two tasks: regulation publication and SAS service state synchronization. First, according to FCC rulings \cite{fcc2015rule}, spectrum regulation for CBRS covers three aspects---incumbent user protection zones, PAL users information and protection zones, and access assignment rules. The G-Chain allows a regulator to distribute such authoritative information to each SAS server.
Second, WInnForum specification \cite{WIForum-SAS} requires that SAS servers synchronize local service states through peer-to-peer communications. The G-Chain can be conveniently instantiated atop the existing framework for inter-SAS server communication and fulfill the task of SAS service state synchronization, via G-Chain transactions.
We define two types of G-Chain transactions:
\begin{itemize}
    \item \textbf{Regulatory Transaction (R-Tx)} is generated by a regulator node indicating the change of spectrum rules, the effect of which can be global or zone-specific.
    \item \textbf{Sync Transaction (S-Tx)} is generated by a SAS server indicating the change of local operational information, such as CBSD registration and coordination events. 
\end{itemize}

All transactions are propagated in the G-Chain network through peer-to-peer gossiping---each SAS server advertises newly received transactions to peers, who then decide whether to fetch the advertised transaction, resembling the transaction propagation model of public cryptocurrency networks.


\textbf{G-Chain Consensus~}
The G-Chain participants need to maintain a unified blockchain ledger that records all occurred transactions through the process of distributed consensus. 
Each participant of the G-Chain consensus (i.e., curators) runs an instance of the consensus protocol which periodically validates and serializes new transactions into blocks.
We note that existing consensus protocols for permissionless and permissioned blockchain networks differ in security threshold, level of decentralization, network scalability, and transaction throughput \cite{vukolic2015quest}. 
In the existing SAS model for CBRS, all SAS servers are managed by administrators with revealed identity, which provides a natural ground for permission control on the G-Chain consensus participants.
With controlled consensus participants, the actual consensus protocol of G-Chain can follow either the proof-of-X (PoX)-Nakamoto style which prioritizes network scalability (commonly used in public blockchains like Bitcoin and Ethereum) or Byzantine fault tolerant (BFT)-style which prioritizes transaction throughput and is mainly used for small-size specialized networks. Hybrid PoX-BFT protocols are also worth considering as they seek a balance between decentralization, network scalability, and transaction throughput \cite{xiao2020survey}. The actual design of G-Chain consensus will need to consider all the factors above and we position it as an open quest. In the convenient case that G-Chain transactions do not require fast finalization as L-Chain transactions do, the consensus protocol of established public blockchains, such as Ethereum's, can be used off-the-shelf.
%

Since G-Chain transactions are designed for information dissemination purposes, transaction validation should focus on the integrity and potentially the correctness of information carried in the transaction payload. 
While transaction integrity can be achieved through public key cryptography, realizing the correctness goal entails extra communication and computation.
For example, when a consensus participant sees a new S-Tx, it may vote valid on the S-Tx if: 1) the majority of the consensus participants who serve that local zone voted valid on the S-Tx, or 2) the participant serves the same zone along with the S-Tx's proposer and recognizes the information in the S-Tx.
On-chain data analytic and statistical inference with past transactions may also provide a ``soft'' determination, which is a potential direction for future work.











\subsection{L-Chain}

An L-Chain for spectrum access assignment is established in each administrative spectrum zone. Participants of an L-Chain include the SAS servers providing service to that zone, local CBSD witnesses, and local CBSD clients. The SAS servers and CBSD witnesses populate the local spectrum committee that curates the L-Chain ledger.
We define three types of L-Chain transactions:
\begin{itemize}
    \item \textbf{Channel Update Transaction (C-Tx)} is issued by a SAS server indicating a change in available channels when the server receives a regulation update from the G-Chain or incumbent appearance notification the ESC.
    \item \textbf{Device Update Transaction (D-Tx)} is issued by a SAS server indicating a CBSD's registration for the local spectrum zone. It replicates partly the CBSD's registration information, including device ID and category, RF configuration onto the L-Chain.
    \item \textbf{Access Request Transaction (A-Tx)} is issued by a CBSD client indicating its request for spectrum access. An A-Tx specifies the sender's device ID, operational parameters (e.g., effective isotropic radiated power (EIRP) and desired frequency range), measurement report, and request type. 
\end{itemize}

A CBSD client registers with a SAS server in compliance with WInnForum's specification \cite{WIForum-CBSD}. 
A C-Tx also allows SAS server to join the L-Chain's committee after a fresh round of server reshuffling (to introduce shortly). 
D-Tx acts as a communication mechanism between the off-chain registration process and L-Chain operation. In addition to replicating new CBSD client's information onto the L-Chain, it also allows a SAS server to mark a CBSD client (in)eligible for the witness status, which subsequently needs approval from the majority of CBSD witnesses.

\textbf{L-Chain Consensus~}
An L-Chain relies on its local spectrum committee, i.e., SAS servers and CBSD witnesses, for transaction settlement and ledger keeping. The inclusion of CBSD witnesses in the L-Chain consensus is designed to get stable users involved in the local spectrum management, who will benefit from an orderly and fault-free spectrum sharing experience.
Meanwhile, L-Chain's responsiveness to spectrum access requests (see Section \ref{subsec:motive-decentsas}) is directly impacted by its consensus efficiency. To speed up L-Chain consensus, we make two complementary design choices that capitalize on the differences between SAS servers and CBSDs.
First, SAS servers and CBSD witnesses are responsible for transaction execution and transaction serialization (i.e., consensus on transaction order in a block) respectively. This is because SAS servers have presumably sufficient computing power for executing transactions but may physically reside far away from the serviced zone; CBSDs are not necessarily powerful computers but physically reside in the zone where they access the spectrum. Close physical proximity yields lower communication delays which result in faster consensus on transaction order.
Second, CBSD witnesses may assume higher mutual trust and thus employ more efficient consensus schemes for themselves, e.g., BFT protocols with lower fault tolerance or crash-fault tolerant (CFT) protocols. 
In Section \ref{subsec:instantiate-with-fabric} we will construct an L-Chain prototype incorporating these designs based on the Hyperledger Fabric platform.

\begin{figure}
    \centering
    \includegraphics[width=0.48\textwidth]{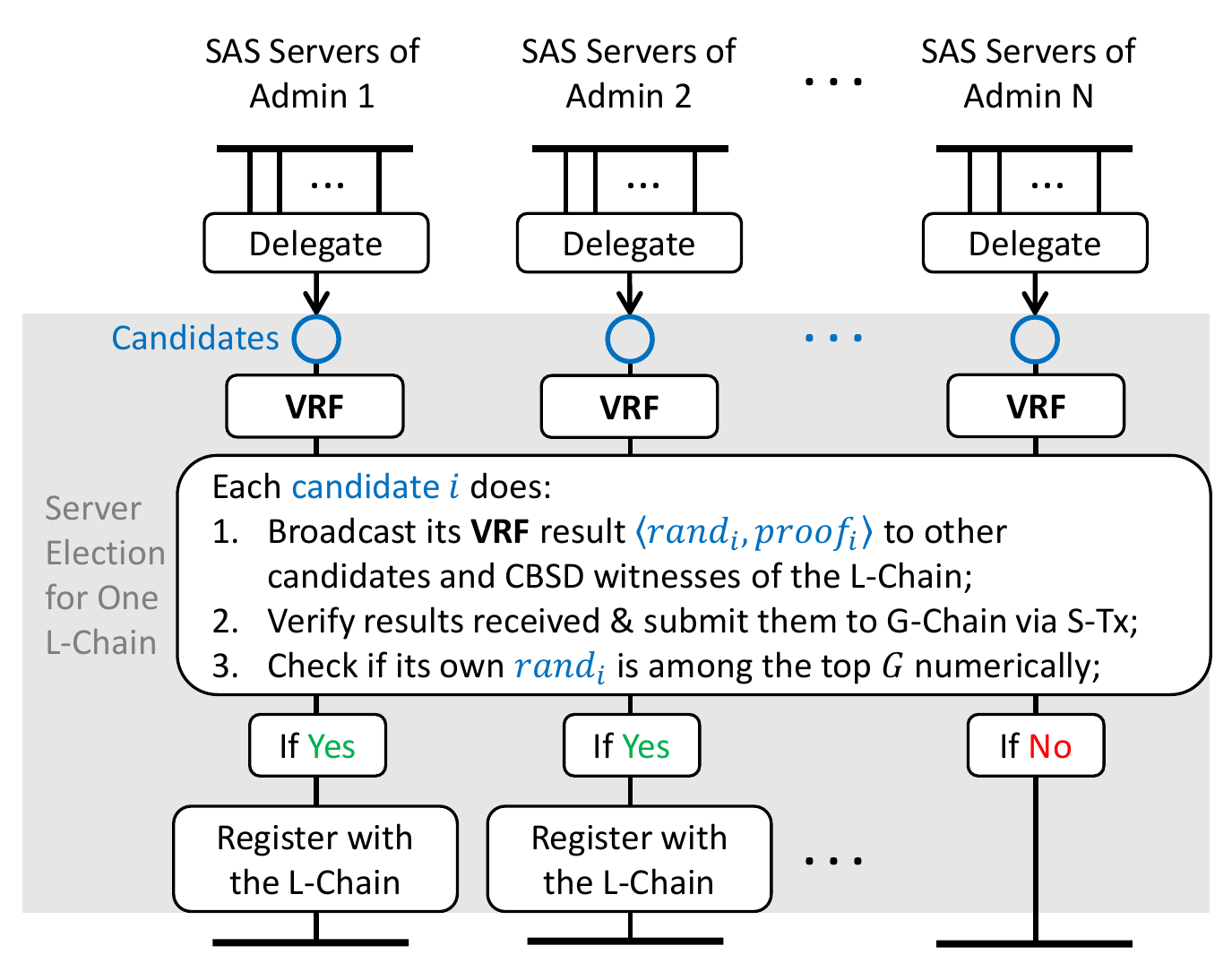}
    \caption{An exemplary SAS server reshuffling procedure in one round for an L-Chain based on verifiable random function (VRF). We assume the SAS server group size for the L-Chain is fixed to $G$.}
    \label{fig:gchain-operation}
\end{figure}

\textbf{SAS Server Reshuffling~}
For a certain period of time, the system needs to undergo a SAS server reshuffling mechanism by which each L-Chain is randomly assigned another round of SAS servers. This offers resilience against an adaptive adversary, who may selectively corrupt individual SAS servers with the aim of compromising a targeted L-Chain.
Here we illustrate an exemplary server reshuffling procedure based on verifiable random function (VRF) in Fig. \ref{fig:gchain-operation}.
The randomness of the VRF output and the verifiability of the output essentially help the L-Chain statistically attain the same fault tolerance threshold as with that on the global level. 
How to realize VRF and the verification mechanism without overloading the G-Chain network stands out as a practical challenge.

\subsection{L-Chain Smart Contract for Spectrum Access Assignment}

Compared to the G-Chain, an L-Chain bears the responsibility of assigning spectrum access to CBSDs. 
A smart contract can be established on top of an L-Chain to automate spectrum assignment at request.
Generally, a smart contract enables the automatic execution of multiparty business logic that involves computation on user-provided data and the transfer of financial tokens. The creation, invocation, and revocation of a smart contract are achieved through blockchain transactions.



\begin{figure}
    \centering
    \includegraphics[width=0.48\textwidth]{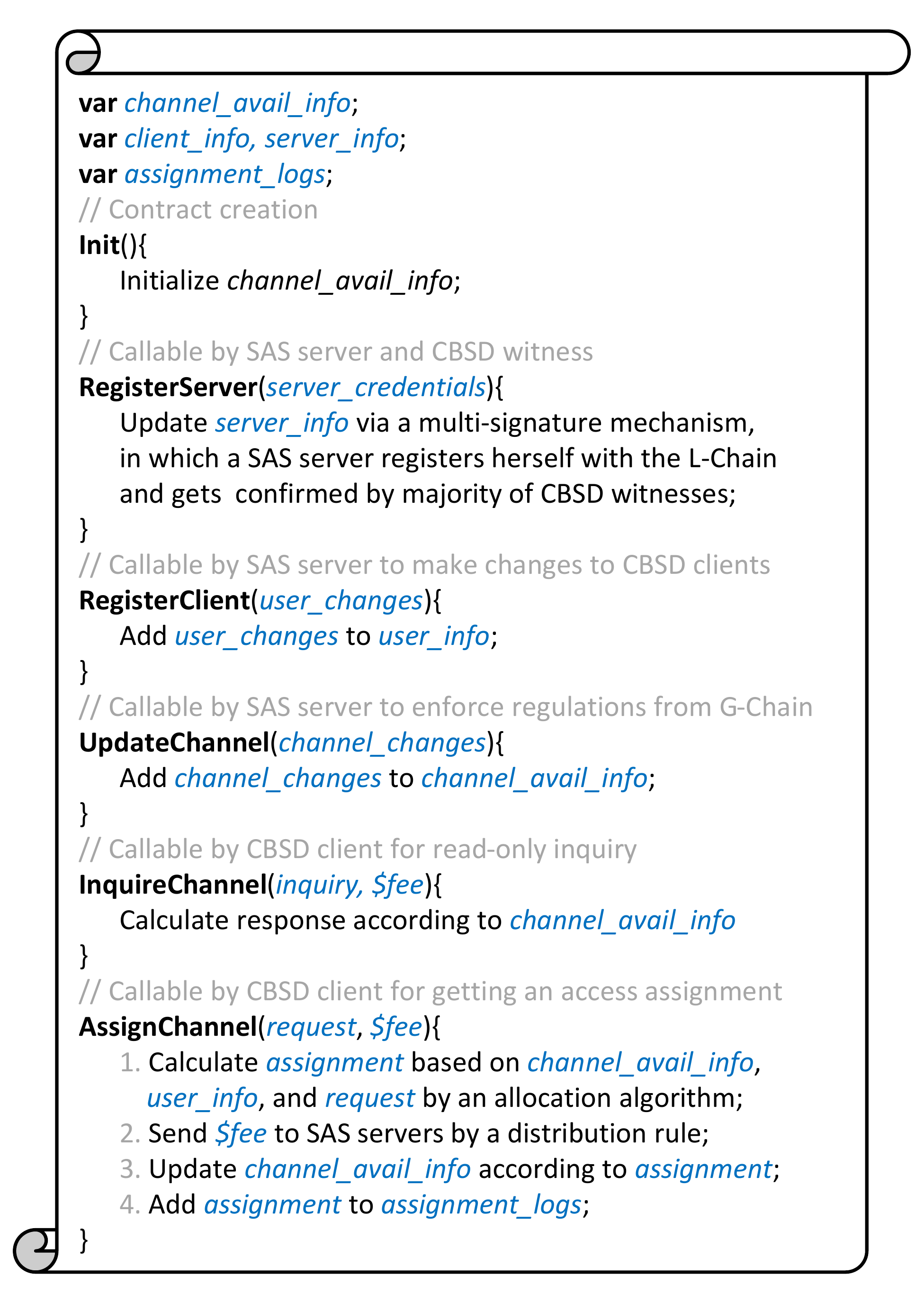}
    \caption{Spectrum access contract $\mathcal{C}_{SA}$ pseudocode.}
    \label{fig:contract}
\end{figure}

\textbf{Encoding Spectrum Access Contract $\mathcal{C}_{SA}$.~}
In compliance with WInnForum's specification on SAS-CBSD interaction \cite{WIForum-CBSD}, the spectrum access contract, denoted $\mathcal{C}_{SA}$, should achieve two goals: 1) enforcing spectrum regulations, and 2) assigning available spectrum bands to spectrum users at request. 
The contract can be either bootstrapped at the onset of L-Chain (i.e., along with the genesis block) or created by SAS servers during operation. The pseudocode of $\mathcal{C}_{SA}$ is shown in Fig. \ref{fig:contract}. To achieve the first goal, a SAS server receiving a notice of incumbent appearance from the ESC or a regulatory update from the G-Chain needs to submit a C-Tx to invoke the \emph{UpdateChannel} function. To achieve the second goal, a SAS server with a newly registered CBSD client needs to submit a D-Tx to invoke the \emph{UpdateClient} to upload the client information. 
A CBSD client needs to submit its spectrum access request in the form of an A-Tx to invoke the \emph{AssignChannel} function. This function encodes an allocation algorithm that calculates an assignment from the existing CBSD information, channel availability, and the request itself. This allocation algorithm should be compliant with regulations including the existing inter-CBSD interference model. The assignment includes the request type (i.e., Grant, Heartbeat, or Relinquish), allocated channel(s), time to expiration, etc. When the A-Tx gets finalized in the L-Chain, the CBSD client can view the assignment on the contract, through a blockchain ledger explorer which can be provided by any SAS server or CBSD witness.
A final consideration is that SAS servers should get compensated for their spectrum access assignment service, as is with the existing SAS' commercial model. 
Accordingly, the \emph{AssignChannel} function is responsible for distributing the client-deposited fee (in the form of G-Chain tokens) to all SAS servers, which will be ultimately cleared in G-Chain by SAS servers via an S-Tx.

\section{Prototyping L-Chain of \sysname}

For spectrum users, their spectrum access experience is most directly impacted by L-Chain performance. In this section, we focus on L-Chain to show the feasibility of our design in terms of performance. We introduce a proof-of-concept instantiation of L-Chain using Hyperledger Fabric and evaluate its performance of spectrum access assignment.
The comprehensive system implementation, including the G-Chain and the server reshuffling mechanism, are deferred to future work.

\subsection{Instantiating L-Chain with Hyperledger Fabric}

Fabric \cite{androulaki2018hyperledger} is a permissioned blockchain framework under the Hyperledger umbrella project started by the Linux Foundation. Comparing to traditional blockchains that follow an ``order-execute'' paradigm wherein transaction serialization and execution are inseparable to the consensus task, Fabric adopts an ``execute-order-validate'' paradigm in that transaction execution and serialization are fulfilled by separate roles. Fabric participants are classified into three roles: \emph{peers} who execute and endorse transaction proposals and validate serialized transactions, \emph{orderers} who provide ordering service to transactions and serialize them into blocks, and \emph{clients} who send transaction proposals to peers, collect endorsements, and send peer-endorsed transactions to orderers for serialization. 
A transaction proposal invokes a \emph{chaincode} (i.e., smart contract) which encodes certain application logic. The \emph{endorsement policy} of the chaincode specifies how many peer endorsements are needed for a transaction to pass validation.


The separation of transaction execution from serialization and the client-in-the-loop endorsement mechanism gives Fabric better architectural modularity, flexibility, and throughput scalability compared to traditional cryptocurrency networks. A production Fabric network can cherry-pick the endorsement policy (for peers) as well as the consensus mechanism (for orderers) according to practical needs on fault tolerance and transaction capacity.
This feature is much appreciated in industrial blockchain applications with stringent delay requirements, such as spectrum access management in our case.

\begin{figure}
    \centering
    \includegraphics[width=0.47\textwidth]{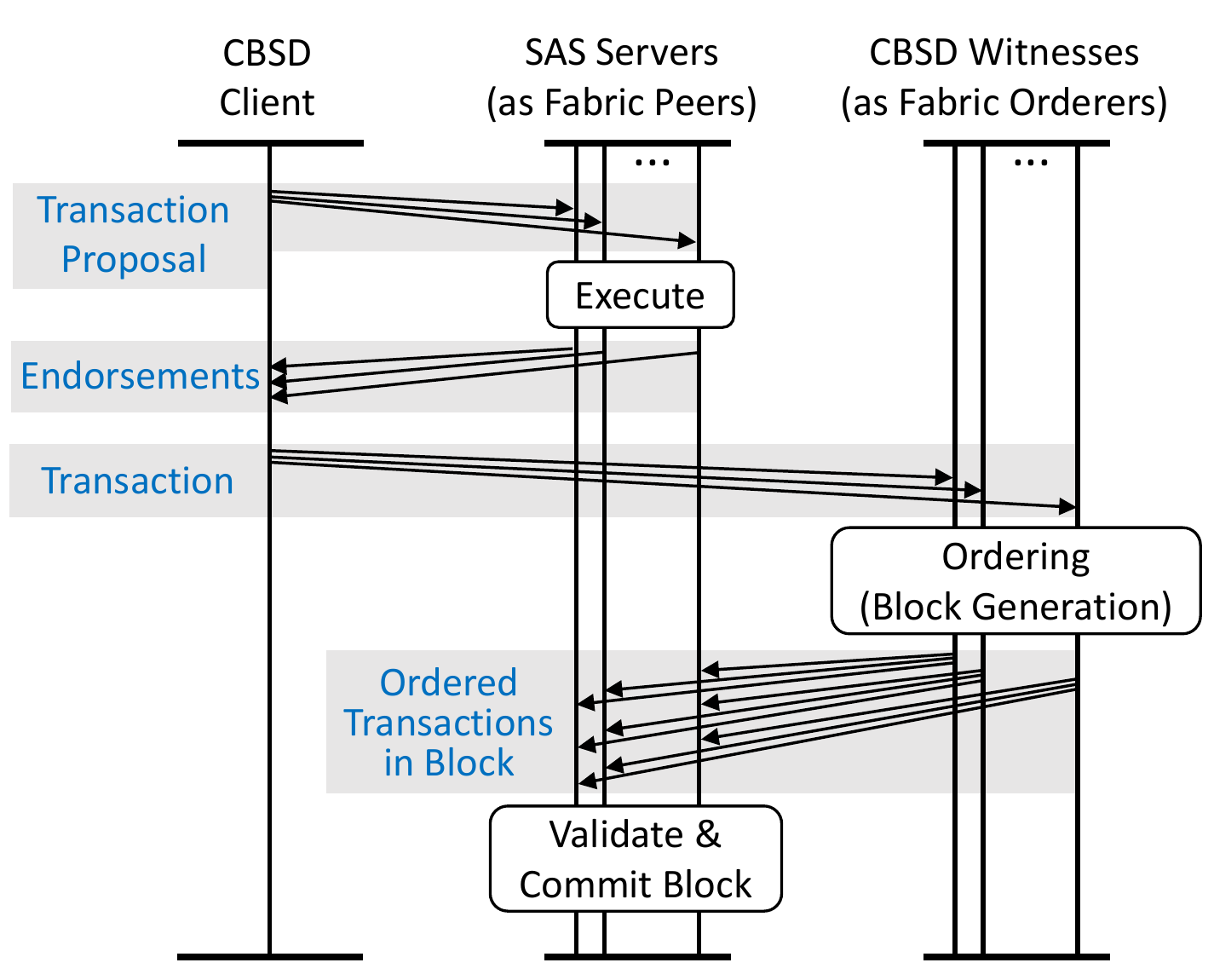}
    \caption{L-Chain transaction processing pipeline based on Hyperledger Fabric's execute-order-validate model \cite{androulaki2018hyperledger}.}
    \label{fig:lchain-operation}
\end{figure}




\textbf{Instantiating L-Chain~}
\label{subsec:instantiate-with-fabric}
In an L-Chain committee, each SAS server operates a Fabric peer and each CBSD witness operates a Fabric orderer. In other words, the SAS servers assume the responsibility of executing transactions while the CBSD witnesses undertake the task of transaction serialization (consensus on the order) and block generation. Fig. \ref{fig:lchain-operation} illustrates the transaction processing pipeline of L-Chain.
The reason behind the task separation is that the local CBSD witnesses have the incentive to provide the ordering service as they are the stable users of the CBRS band and will directly benefit from the safe operation of the L-Chain.
We opt for Fabric's native Raft consensus, an efficient CFT consensus protocol, for the ordering service. In this evaluation we do not assume Byzantine CBSD witnesses.

\subsection{Evaluating Spectrum Access Assignment}
We implemented three L-Chain prototypes, each consisting of $4$ SAS servers and $5/10/20$ CBSD witnesses. For each prototype, the SAS servers and CBSD witnesses were deployed in docker containers as a virtual network in a Linux machine with 4 cores and 16GB memory. The L-Chain block interval was fixed to $1$ second and each block could include up to $1$MB of transactions. The endorsement policy was set to 3 out of 4 peers.
Each prototype emulated a simple L-Chain network and was purposed for demonstrating the feasibility of our Fabric-based design. 
Our evaluation focused on the core task of spectrum access assignment (i.e., client sending out A-Tx to invoke the \emph{AssignChannel} function of $\mathcal{C}_{SA}$), as it most directly impacts \sysname' quality of service. Specifically, two metrics were evaluated: \emph{transaction finalization latency} of A-Tx, which is the time between when a CBSD client submits an A-Tx and when the assignment gets finalized in the L-Chain ledger, and \emph{throughput}, which measures the maximum A-Tx transactions per second (TPS) the L-Chain can handle without be overloaded.

\begin{table}
    \centering
    \caption{Spectrum Access Assignment Performance (clients send A-Tx to call $\mathcal{C}_{SA}$'s \emph{AssignChannel}) of an L-Chain with 20 CBSD witnesses under four delay regimes.}
    \begin{tabular}{c|c|c|c|c|c}
    \toprule
         & Packet Delay & \multicolumn{3}{c|}{Tx Finalization Latency} &  Throughput \\
         & Mean / Jitter & Min & Mean & Max & (TPS) \\
    \midrule
        1 & 10ms / 1ms & 0.42s & 1.06s & 1.87s & 61.1 \\
        2 & 30ms / 3ms & 0.73s & 1.47s & 2.36s & 54.2 \\
        3 & 50ms / 5ms & 1.09s & 1.93s & 2.73s & 48.4 \\
        4 & 100ms / 10ms & 1.36s & 2.99s & 3.63s & 31.2 \\
    \bottomrule
    \end{tabular}
    \label{tab:result-assignment-delay}
\end{table}

\begin{figure}
    \centering
    \vspace{-.15in}
    \hspace{-.2in}
    \subfigure[Throughput v.s. CBSD witnesses]{
        \centering
        \includegraphics[width=0.234\textwidth]{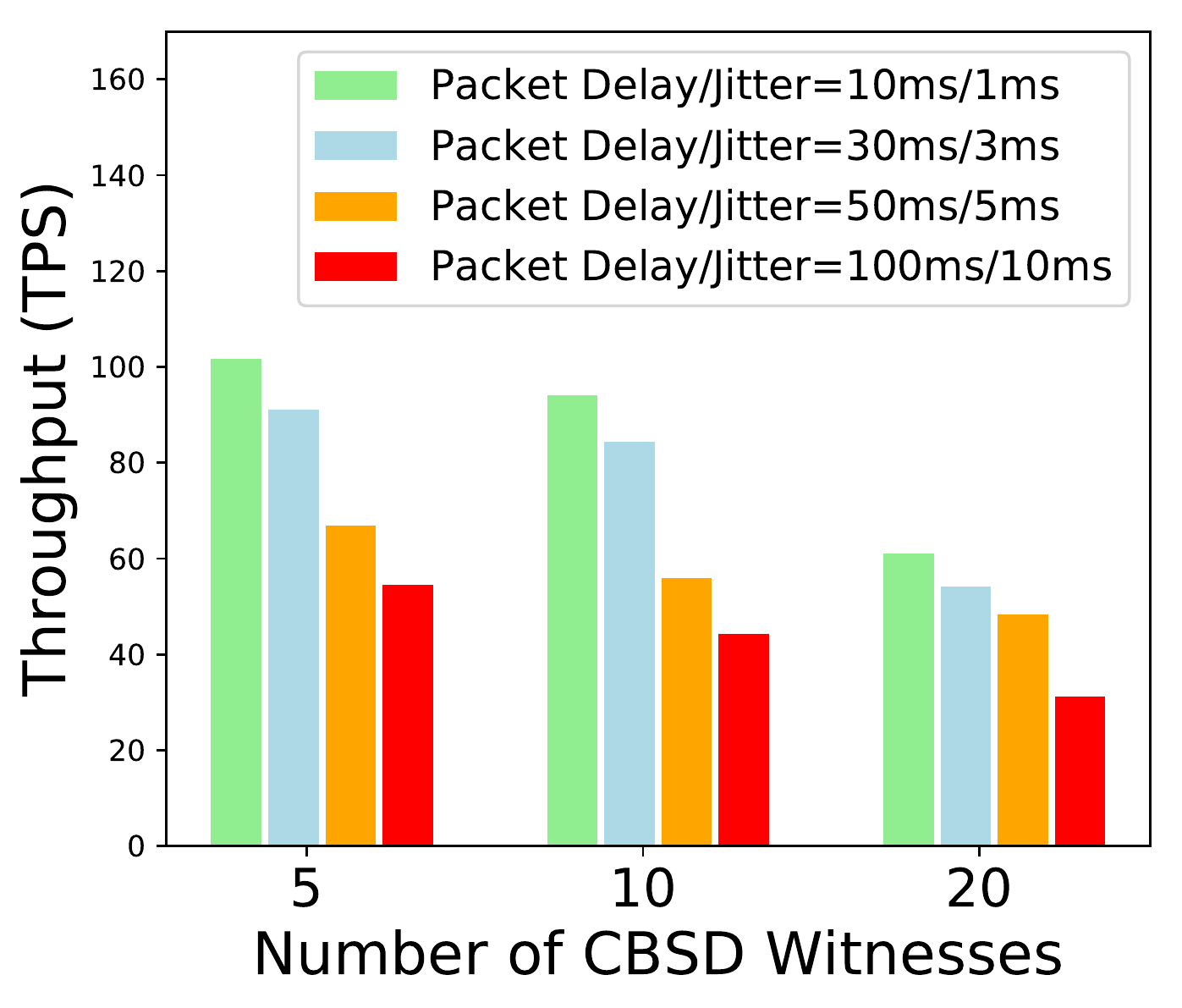}
        \label{fig:l-chain-performance-a}
    }
    \hspace{-.1in}
    \subfigure[Latency v.s. throughput]{
        \centering
        \includegraphics[width=0.24\textwidth]{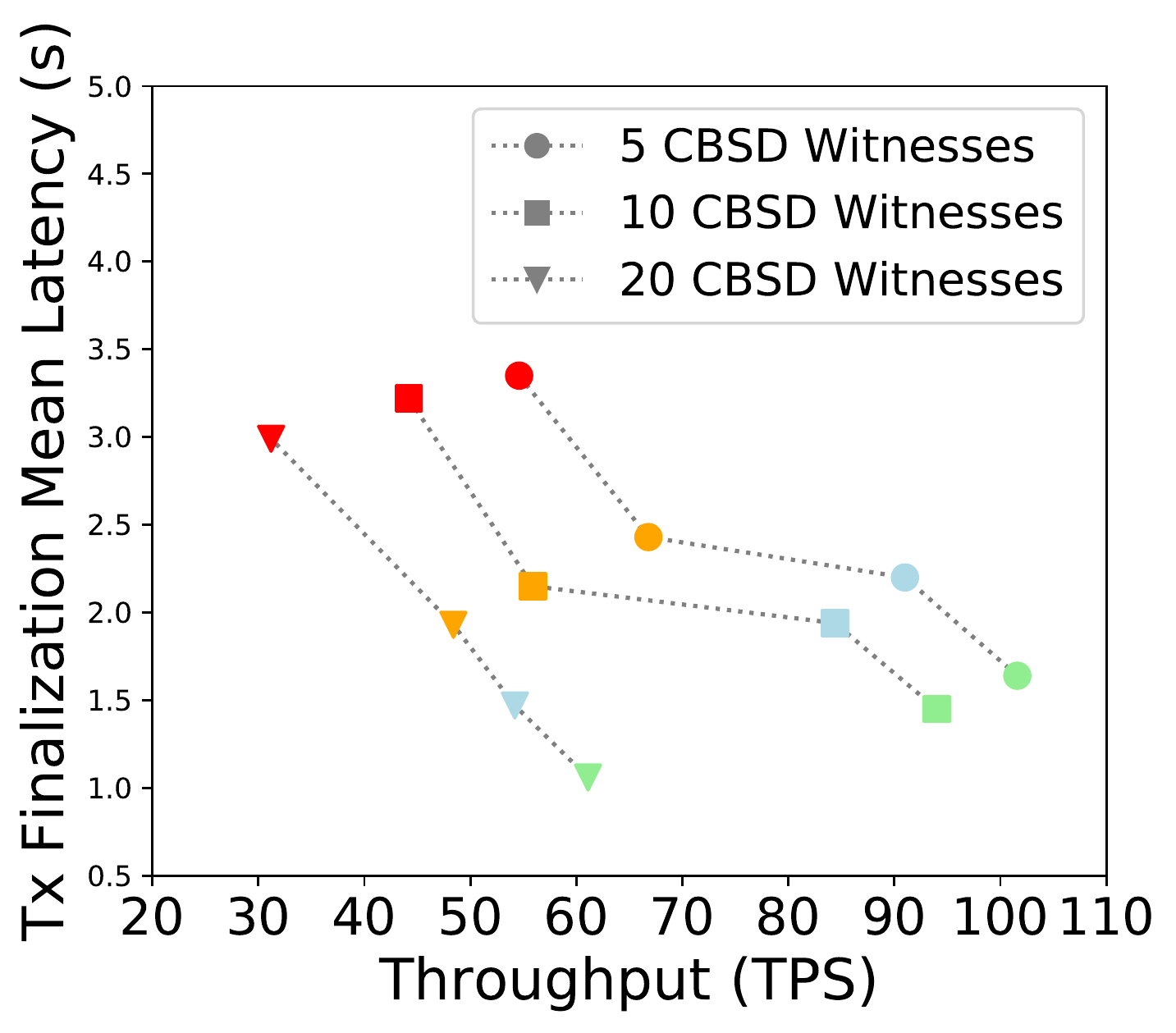}
        \label{fig:l-chain-performance-b}
    }
    \hspace{-.1in}
    \caption{Spectrum access assignment performance (clients send A-Tx to call $\mathcal{C}_{SA}$'s \emph{AssignChannel}) of three L-Chains under four delay regimes. }
    \label{fig:l-chain-performance}
\end{figure}

We used Hyperledger Caliber\footnote{https://hyperledger.github.io/caliper}, a blockchain benchmarking tool, to simulate a constant A-Tx traffic invoking \emph{AssignChannel}, from $50$ CBSD clients. To simulate network delays in the Internet, we added packet delays to every entity using the Linux traffic control tool from the iproute2 package\footnote{https://wiki.linuxfoundation.org/networking/iproute2}. The four delay regimes (mean/jitter) are specified in Table \ref{tab:result-assignment-delay}. Table \ref{tab:result-assignment-delay} shows the performance result for the L-Chain prototype with 20 CBSD witnesses. Under the harshest delay regime ($100ms/10ms$), the system is able to finalize an A-Tx within five seconds. In CBRS operation, this short finalization latency would allow a CBSD client to perform the Heartbeat procedure gracefully, as the Heartbeat interval is typically set at 30 or 60 seconds \cite{google2020sasresponse}.
On the other hand, transaction throughput decreases with longer packet delays or when the CBSD witness population increases, as is shown in Fig. \ref{fig:l-chain-performance-a}. Fig. \ref{fig:l-chain-performance-b} shows the trade-off between transaction finalization latency and throughput, with the four delay regimes and CBSD witness populations. Both results are inline with the prior wisdom that a permissioned blockchain system should have good inter-node connectivity and limited network size in order to attain high throughput and low finalization latency \cite{xiao2020survey}. We note that there are other degrees of freedom for performance optimization, such as using a more efficient consensus design and fine-tuned block interval, which we will explore in future implementations.

\section{Broader Challenges and Opportunities}

The \sysname framework provides a starting point for designing future decentralized spectrum management systems which may incorporate more sophisticated user requirements across wider frequency bands. Here we identify important challenges and opportunities for \sysname and blockchain-based SAS solutions.

\textbf{Privacy Protection for Spectrum Users:~} 
Spectrum access assignment of \sysname, which is executed by an L-Chain in a transparent manner, may risk exposing sensitive operational information of CBSD clients. This poses a privacy disadvantage for \sysname compared to the existing SAS with an end-to-end service model. There are two directions to protect spectrum user privacy in \sysname and blockchain-based schemes.
First, user information can be obfuscated in the blockchain ledger. SAS servers use an obfuscated user operational information when it submits a D-Tx, however the obfuscation is only applied to the extent that the result remains useful for the assignment task.
Second, hardware-assisted trusted computing techniques, such as trusted execution environment (TEE), can be instantiated alongside the blockchain instance in a SAS server. Spectrum user information publicized on a smart contract can be encrypted by the user and the execution of spectrum assignment routines can be offloaded to a secure TEE container. User information is decrypted and used for computation in the container after the user provides its decryption key through an attested secure channel.
Challenges remain on the secure commitment of assignment results from the container to the blockchain \cite{xiao2020privacyguard} and the security of TEE technology which depends on the hardware vendor.



\textbf{Enabling Secondary Spectrum Market:~}  
In a 2016 revision of its CBRS ruling \cite{fcc2015rule} the FCC suggested the establishment of secondary markets for trading spectrum usage rights held by PAL licensees. The vision was that the market mechanism would lead to more efficient and dynamic spectrum usage. In the centralized spectrum management model, a trusted spectrum exchange can be established for matching buy and sell orders from license holders and potential buyers \cite{caicedo2011viability}. In comparison, a blockchain-based decentralized spectrum exchange would have an advantage in transparency and decentralization, which provides higher robustness when no centralized exchanges can be relied upon.
With the establishment of a smart contract environment and a native transactional framework, spectrum usage rights (of a certain band, locality, time) as well as derivatives can be publicly traded, facilitated by blockchain transactions from buyers and sellers. 
One major challenge lies in the ``securitization'' of spectrum access rights and the enforcement of trades, since wireless spectrum, unlike ordinary financial or physical assets, is self-existent and intangible. 
Another challenge remains in realizing the business model of a blockchain-based spectrum exchange, including trader incentives, orders matching, commission fee assessment \cite{weiss2019application,shi2021challenges}.
To address these challenges, we expect a multidisciplinary approach involving blockchain, finance, and wireless communications.

\textbf{Extending to Wider Bands and Heterogeneous Spectrum Sharing Models:~}
A future spectrum sharing system is expected to manage spectrum access for a large variety of applications and users which need higher automation, adaptability, and finer granularity. The SAS concept may extend beyond the 3.5 GHz CBRS band and be applied to heterogeneous spectrum sharing models. For unlicensed spectrum such as the 6GHz band, a blockchain-based SAS can potentially bring security and incentive for honest participation to an opportunistic spectrum sharing landscape. For non-commercial users, such as astronomical observatories who need to observe wide spectrum bands for scientific discovery and weather radars which need to actively transmit exploratory signals, a blockchain-based SAS can provide a trustworthy bulletin board for them to publicize their observation/exploration schedules and serve as a marketplace for trading underused bands.
Amid such heterogeneity of spectrum users and service requirements, how to ensure secure exchange of sensitive information with an appropriate level of privacy will be a key research challenge.

\section{Conclusion}
A decentralized SAS framework poses a promising solution for robust and fault-tolerant spectrum access management. In this paper, we raised the practical requirements and proposed a blockchain-based decentralized SAS architecture called \sysname. \sysname consists of two layers: a G-Chain for SAS service state synchronization at the global scale and L-Chains which provide spectrum access management services to spectrum users at the local scale, without having the users trust individual SAS servers. We implemented a \sysname prototype with Hyperledger Fabric and evaluated its spectrum assignment performance in an emulated L-Chain network.
We concluded with broader challenges and opportunities for \sysname and general blockchain-based SAS solutions.


\section*{ACKNOWLEDGMENT}
This work was supported in part by the U.S. National Science Foundation under Grant AST-2037777, Grant CNS-1916902, Grant CNS-1916926; in part by the Virginia Commonwealth Cyber Initiative (CCI); and a gift from InterDigital.

\end{document}